\documentclass[%
 aip,
 jmp,%
 amsmath,amssymb,
 reprint,%
]{revtex4-1}

\usepackage{graphicx}
\usepackage{dcolumn}
\usepackage{bm}

\begin{document}

\preprint{AIP/123-QED}

\title{Optimal and suboptimal networks for efficient navigation measured by mean-first passage time of random walks}

\author{Zhongzhi Zhang}
\email{zhangzz@fudan.edu.cn}
\homepage{http://homepage.fudan.edu.cn/~zhangzz/}

\affiliation {School of Computer Science, Fudan University, Shanghai
200433, China}

\affiliation{Shanghai Key Lab of Intelligent Information
Processing, Fudan University, Shanghai 200433, China}

\affiliation{State Key Laboratory for Novel Software Technology, Nanjing University, Nanjing 210023, China}

\author{Yibin Sheng}

\affiliation{School of Mathematical Sciences, Fudan University,
Shanghai 200433, China}

\author{Zhengyi Hu}

\affiliation {School of Computer Science, Fudan University, Shanghai
200433, China}

\author{Guanrong Chen}
\affiliation {Department of Electronic Engineering, City University of Hong Kong, Hong Kong SAR, China}

\date{\today}

\begin{abstract}
For a random walk on a network, the mean first-passage time from a node $i$ to another node $j$ chosen stochastically according to the equilibrium distribution of Markov chain representing the random walk is called Kemeny constant, which is closely related to the navigability on the network. Thus, the configuration of a network that provides optimal or suboptimal navigation efficiency is a question of interest. It has been proved that complete graphs have the exact minimum Kemeny constant over all graphs. In this paper, by using another method we first prove that complete graphs are the optimal networks with a minimum Kemeny constant, which grows linearly with the network size. Then, we study the Kemeny constant of a class of sparse networks that exhibit remarkable scale-free and fractal features as observed in many real-life networks, which cannot be described by complete graphs. To this end, we determine the closed-form solutions to all eigenvalues and their degeneracies of the networks. Employing these eigenvalues, we derive the exact solution to the Kemeny constant, which also behaves linearly with the network size for some particular cases of networks. We further use the eigenvalue spectra to determine the number of spanning trees in the networks under consideration, which is in complete agreement with previously reported results. Our work demonstrates that scale-free and fractal properties are favorable for efficient navigation, which could be considered when designing networks with high navigation efficiency.
\end{abstract}


\maketitle

\begin{quotation}
Kemeny constant, defined as the mean first-passage time from a node $i$ to a node $j$ selected randomly according to the equilibrium distribution of Markov chain, is a fundamental quantity for a random walk, since it is a useful indicator characterizing the efficiency of navigation on networks. Here we prove that among all networks, complete graph is the optimal network having the least Kemeny constant, which is consistent with the previous result obtained by a different method. Then, we study the Kemeny constant of a class of sparse scale-free fractal networks, which are ubiquitous in real-life systems. By using the renormalization group technique, we derive the explicit formulas for all eigenvalues and their multiplicities of the networks, based on which we determine the closed-form solution to the Kemeny constant. We show that for some particular cases of networks the leading scaling of Kemeny constant displays the same behavior as that of complete graph, indicating that for networks with many nodes, navigation performance similar to that of complete graph can be obtained by networks with many few links. Finally, to show the validity of the eigenvalues and their degeneracies, we also use them to count spanning trees in the networks being studied, and recover previously reported results. Our work is helpful for the structure design of networks where navigation is very efficient.
\end{quotation}

\section{Introduction}

In the past decade, many research reports demonstrated that complex networks are ubiquitously present in nature and social sciences~\cite{AlBa02,Ne03,BoLaMoChHw06}. Today, complex networks have been recognized as an emerging subject and useful tool for studying complex systems~\cite{Ne10}. An outstanding issue in network science is to understand the behavior of various dynamical and stochastic processes taking place on different networks with rich topologies, which is not only of theoretical interest but also of practical relevance~\cite{Ne03,DoGoMe08}. Among diverse dynamics, random walks~\cite{NoRi04,CoBeTeVoKl07,ZhAlHoZhCh11} are a paradigmatic process and have been extensively studied~\cite{SoRebe05,Bobe05,GaSoHaMa07,ZhLiZhWuGu09,FrFr09,ZhZhZhYiGu09,RoHa11,TeBeVo11,CaBeIvCa12} due to their wide range of applications in science and engineering~\cite{We94,Hu95,FoPiReSa07,BeLoMoVo11}.

A fundamental quantity for random walks is mean first-passage time (MFPT)~\cite{Re01}. The MFPT from a source node $i$ to a destination node $j$, denoted by $F_{ij}$, is the expected time for a walker starting from $i$ to reach $j$ for the first time. Many other quantities related to random walks can be expressed in terms of, or encoded in, MFPT. For example, for the trapping problem~\cite{Mo69}---a particular case of random walks with a trap located at a given node absorbing walkers visiting it---the trapping efficiency quantified by average trapping time is defined as the mean of node-to-trap MFPT over source nodes. As a quantitative indicator of trapping efficiency, average trapping time has been studied on various networks with distinct topological properties, with an attempt to uncover the impact of network topology on the trapping process. Examples include classic fractals~\cite{KaBa02PRE,KaBa02IJBC,Ag08,HaRo08,LiWuZh10,ZhWuCh11}, non-fractal scale-free networks~\cite{KiCaHaAr08,ZhQiZhXiGu09,ZhGuXiQiZh09,AgBu09,AgBuMa10,MeAgBeVo12}, fractal scale-free networks~\cite{ZhXiZhGaGu09,TeBeVo09,ZhYaG11}, modular scale-free networks~\cite{ZhLiGaZhGuLi09,ZhYaLi12}, and so on.

In addition to the trapping problem, another interesting quantity derived from MFPT is the  Kemeny constant, which is defined as the expected time for a walker starting from a node $i$ to another node $j$ selected randomly from all nodes according to the stationary distribution of Markov chain representing the random walk~\cite{KeSn76}. Since the Kemeny constant is equal to the sum of reciprocals of one minus each eigenvalue of the  transition matrix other than eigenvalue one~\cite{Lo96}, it is also called eigentime identity~\cite{AlFi99}, which is independent of the starting point. It can be explained in terms of navigability~\cite{LeLo02}, measuring the efficiency of navigation. Thus, it is of theoretical and practical significance to study the Kemeny constant of complex networks. In contrast to the average trapping time, little attention has been paid to the Kemeny constant, and many questions about this relevant quantity are still open.

In this paper, we address a specific question: of all networks, which have a minimum or almost minimum Kemeny constant? It is of theoretical relevance and practical importance, and is particularly useful for network design. By making use of the connection between the Kemeny constant and eigenvalues of the transition matrix, we first show that among all undirected networks complete graphs have the minimum Kemeny constant growing linearly with the network size, which agrees with result reported before~\cite{AlFi99}. However, since most real networks are spare in the sense that their average degrees are much less than that of complete graphs. Moreover, it was shown that many real networks exhibit the striking scale-free~\cite{BaAl99} and fractal~\cite{SoHaMa05} properties. Here we study the Kemeny constant for a class of sparse scale-free fractal networks~\cite{ZhZhZo07}, denoted by $H(q,n)$, which are parameterized by a positive integer $q \geq 2$. Using the decimation technique, we derive explicit expressions for all eigenvalues and their multiplicities of the transition matrix of $H(q,n)$, based on which we derive a closed-form formula for the Kemeny constant. The obtained result shows that, for $q \geq 3$, the Kemeny constant also scales linearly with the network size, displaying the same scaling as that of complete graphs. Finally, we use the eigenvalues of $H(q,n)$ to determine the number of spanning trees and recover previous results, which also validates our computation of eigenvalues for such networks.

\section{Formulation of the problem \label{RanWalk}}

Consider discrete-time unbiased random walks~\cite{NoRi04} on a connected undirected network $G$ with $N$
nodes and $E$ edges, where the $N$ nodes are labeled by $1,2,3,\ldots, N$, respectively. The connectivity of $G$ is
encoded in its adjacency matrix $A$, whose entry $a_{ij}=1$ (or 0) if nodes $i$ and $j$ are (not) connected by an edge. The
degree of node $i$ is defined to be $d_i=\sum_{j=1}^{N} a_{ij}$, and the diagonal degree matrix of $G$, denoted by $D$, is defined
as: the $i$th diagonal element is $d_i$, while all non-diagonal elements are zero. Thus, the total degree of all
nodes is $K=2E= \sum_{j=1}^{N}d_i$, and the average degree is $\langle d \rangle=2E /N$.

For the unbiased random walks considered here, at every time step the walker starting from its current location jumps to each of its neighbors with an identical probability. Such a stochastic process  is characterized by the transition matrix $T=D^{-1}A$, whose entry $t_{ij}= a_{ij}/d_i$ presents the probability of moving from $i$ to $j$ in one step. Note that the transition matrix is also called the Markov matrix, since a random walk is an ergodic Markov chain~\cite{KeSn76,AlFi99}, whose stationary distribution $\pi=(\pi_1, \pi_2,\ldots, \pi_N)^\top$ is a unique probabilistic vector satisfying $\pi_i=d_i/K$, $\sum_{i=1}^{N}\pi_i=1$, and $\pi^{\top}T=\pi^{\top}$.

In general, the transition matrix $T$ is asymmetric except for regular networks. So, we introduce a matrix $P$ similar to $T$  by
\begin{equation}\label{Trans01}
P=D^{-\frac{1}{2}} A D^{-\frac{1}{2}}=D^{\frac{1}{2}} T D^{-\frac{1}{2}},
\end{equation}
where $D^{-\frac{1}{2}}$ is defined as follows~\cite{Ch97}: the $i$th diagonal entry is $1/\sqrt{d_i}$, while all non-diagonal entries are equal to zero. It is evident that $P$ is real and symmetrical and thus has the same eigenvalues as $T$.

Various interesting quantities related to random walks are encoded in matrix $P$, and thus can be deduced from $P$. For example, the MFPT $F_{ij}$ for a walker starting from a node $i$ to another node $j$ can be expressed in terms of the eigenvalues and their orthonormalized eigenvectors of matrix $P$. Let $\lambda_1$, $\lambda_2$, $\lambda_3$, $\ldots$, $\lambda_N$ be the $N$ (real) eigenvalues of matrix $P$, rearranged as $1=\lambda_1>\lambda_2 \geq \lambda_3 \geq \cdots \geq \lambda_N \geq -1$. Obviously, $\lambda_1+\lambda_2 + \lambda_3 + \cdots + \lambda_N =0$. And let $\psi_1$, $\psi_2$, $\psi_3$, $\ldots$, $\psi_N$ denote the corresponding normalized, real-valued and mutually orthogonal eigenvectors, where $\psi_i=(\psi_{i1},\psi_{i2},\ldots,\psi_{iN})^{\top}$. Then,
\begin{equation}\label{MFPT01}
F_{ij}=\frac{K}{d_{j}}\sum_{k=2}^{N}\frac{1}{1-\lambda_{k}}\left(\psi_{kj}^{2}-\psi_{ki}\psi_{kj}\sqrt{\frac{d_{j}}{d_{i}}}\right)\,,
\end{equation}
which has been derived by various methods, e.g., generating functions~\cite{ZhAlHoZhCh11} and spectral graph theory~\cite{Lo96}.

Another interesting quantity for random walks is the Kemeny constant~\cite{KeSn76}, also referred to as eigentime identity~\cite{AlFi99}, defined by
\begin{equation}\label{Kemeny01}
F =\sum_{j=1}^{N}\pi_j\,F_{ij}\,.
\end{equation}
Equation~(\ref{Kemeny01}) indicates that $F$ is the average of MFPT $F_{ij}$ from node $i$ to  node $j$ randomly chosen from all nodes accordingly to the stationary distribution.  Since $F$ is independent of the starting point $i$, it is thus called the Kemeny constant. This quantity is closely related to user navigation through World Wide Web and can be explained as the mean number of edges the random surfer needs to follow before reaching his/her final destination~\cite{LeLo02}.

Note that the Kemeny constant is a global spectral characteristic of a
network~\cite{KeSn76,AlFi99,ZhAlHoZhCh11}. It has been shown~\cite{KeSn76,AlFi99,ZhAlHoZhCh11}  that
\begin{equation}\label{Kemeny01}
F =\sum_{j=1}^{N}\pi_j\,F_{ij} \equiv  \sum_{k=2}^{N}\frac{1}{1-\lambda_{k}}\,,
\end{equation}
which holds for any connected undirected network. Kemeny constant can be used as a measure of navigation efficiency~\cite{LeLo02} and is thus relevant to Web design: small $F$ can save user's effort by directly guiding them to most desirable Web pages. Since $F$ depends on the network structure, a question arises naturally: what is the optimal network structure that has a minimum Kemeny constant? Intuitively, the globally optimal network should be fully connected.
In the sequel, we will address this question analytically by using the spectra of graphs and method of Lagrange multipliers, and show that the eigenvalue treatment arrives at the intuitive and natural conclusion.

\section{Globally optimal network \label{Optimal}}

We now solve the aforementioned problem by using the method of Lagrange multipliers. We consider the following function:
\begin{equation}\label{Optima01}
f(x_2,\cdots,x_N) = \sum_{k=2}^{N}\frac{1}{1-x_{k}}\,,
\end{equation}
subject to the condition of $1+x_2+\cdots +x_N=0$ on the domain $1>x_2 \geq x_3 \geq \cdots \geq x_N \geq -1$. Applying the method of Lagrange multipliers to Eq.~(\ref{Optima01}), the minimum of the function $f(x_2,\cdots,x_N)$ is attained at $x_2 = x_3 = \cdots = x_N = -1/(N-1)$. Thus, the minimum Kemeny constant among all networks of size $N$ is
\begin{equation}\label{Optima02}
F_{\rm min} =\frac{(N-1)^2}{N}\,.
\end{equation}

As one can see, the above set of identical eigenvalues $\lambda_2 = \lambda_3 = \ldots = \lambda_N =-1/(N-1)$ corresponds to those of the transition matrix of a complete graph with $N$ nodes, in which every node is connected to all others. That is, among all connected undirected networks, complete network is optimal in the sense that it has the absolute minimum Kemeny constant, which can not be achieved by other networks. We note Eq.~(\ref{Optima02}) has been previously proved by using a different approach~\cite{AlFi99}. It is obvious from Eq.~(\ref{Optima02}) that for large $N$, $F_{\rm min}$ scales linearly with $N$.

Although a complete network has the minimum  Kemeny constant, it is dense, not like most real networks. Extensive empirical research has shown~\cite{AlBa02,Ne03,BoLaMoChHw06} that most real networks are sparse, which have a small average degree and display simultaneously scale-free~\cite{BaAl99} and fractal~\cite{SoHaMa05} properties. Below, we will study the Kemeny constant for a family of fractal scale-free sparse networks~\cite{ZhZhZo07} parameterized by a positive integer $q \geq 2$. We will show that for those networks corresponding to $q \geq 3$, the dominating term of their Kemeny constant grows linearly with the network size $N$, which is similar to the linear behavior of complete graphs but with a different prefactor $\frac{q-1}{q-2}$ larger than $1$. Thus, when $q \geq 3$, the networks are called  ``suboptimal networks''.

\begin{figure}
\begin{center}
\includegraphics[width=0.75\linewidth]{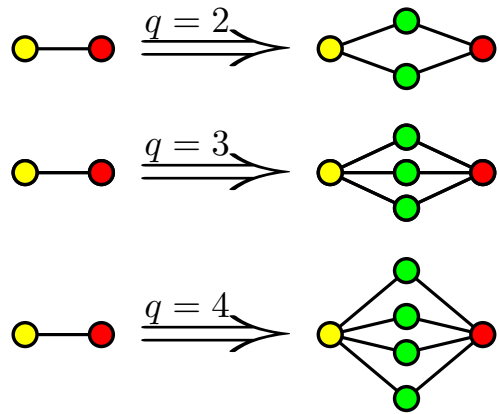}
\caption{(Color online) Construction  of the
networks. One can obtain the next generation of the network family through replacing each edge of the present generation by the clusters on the right-hand side of the arrows.}
\label{cons}
\end{center}
\end{figure}

\section{Suboptimal scale-free fractal networks \label{Suboptimal}}

In the section, we study random walks on a class of scale-free (heterogenous) fractal networks~\cite{ZhZhZo07}, the properties of which are dominated by an integer parameter $q \geq 2$.

\subsection{Construction and properties \label{model}}

\begin{figure}
\begin{center}
\includegraphics[width=1.0\linewidth]{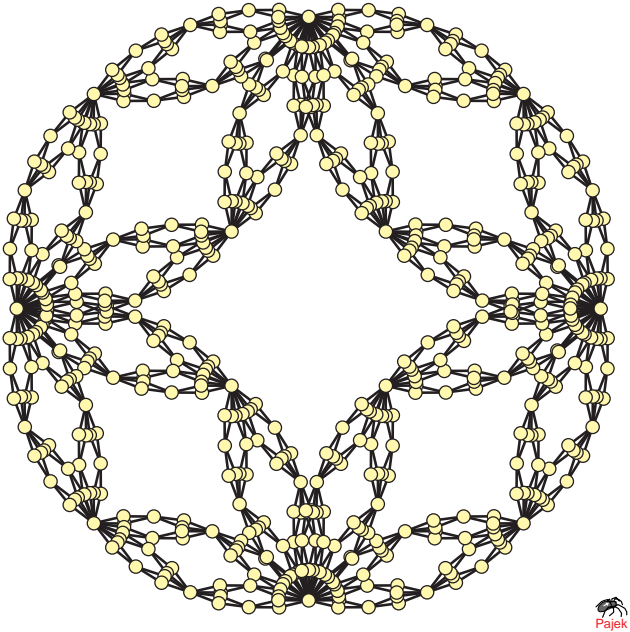}
\end{center}
\caption[kurzform]{ (Color online) Illustration for the network $H(2,5)$.} \label{network}
\end{figure}

The scale-free fractal networks studied in~\cite{ZhZhZo07} are constructed in an iterative manner as depicted by Fig.~\ref{cons}. Let $H(q,n)$ ($q\geq 2$ and $n\geq 0$) denote the networks after $n$ iterations. Initially ($n=0$), $H(q,0)$ is an edge connecting two nodes. For $n\geq 1$, $H(q,n)$ is generated from $H(q,n-1)$ by replacing every existing edge in $H(q,n-1)$ with the clusters on the right-hand side of the arrows in Fig.~\ref{cons}. When $q=2$, $H(q,n)$ is reduced to the $(2,2)-$flower, which is a special case of the $(u,v)-$flowers ($u \geq 1$, $v \geq 2$)  presented in~\cite{RoHaAv07,RoAv07}. Figure~\ref{network} illustrates the network $H(2,5)$ for the particular case of $q=2$, which is  called diamond lattice and was introduced more than thirty years ago~\cite{BeOs79}.

The iterative construction allows to precisely analyze relevant properties of the networks. At each generation $n_i$ ($n_i\geq 1$), the number of newly introduced nodes is $V_{n_i}=q(2q)^{n_i-1}$. Then, the network size (number of nodes) $N_n$ of $H(q,n)$ is
\begin{equation}\label{Nn}
N_n=\sum_{n_i=0}^{n}V_{n_i}=\frac{q(2q)^{n}+3q-2}{2q-1}\,.
\end{equation}
Let $E_n$ denote the number of total edges in $H(q,n)$. It is easy to verify that
\begin{equation}\label{En}
E_n=qE_{n-1}=(2q)^{n}\,.
\end{equation}
Let $d_i(n)$ be the degree of  node $i$ in $H(q,n)$ that was generated at iteration $n_i$ ($n_i\geq 1$). Then, we have $d_i(n+1)=q\,d_i(n)=2\,q^{n-n_i}$.

These resultant networks display some remarkable features observed in various real-life systems. They have a power-law degree distribution with exponent $\gamma=2+\ln 2/ \ln q$, and have a fractal dimension $f_{\rm B}=\ln (2q)/ \ln 2$, both of which show that they are fractal scale-free networks~\cite{ZhZhZo07,RoHaAv07,RoAv07}. Moreover, they are ``large-world'', with both of their diameter and average distance increasing as a power function of the network size~\cite{ZhZhZo07,RoHaAv07,RoAv07}. We note that this ``large-world'' phenomenon was also observed in the global network of avian influenza outbreaks~\cite{SmWaTs07,SmXuZhZhSuLu08}.

\subsection{Spectrum of the transition matrix}

After introducing the construction method and properties of the networks, we proceed to determine the eigenvalues and their degeneracies of the transition matrix for the network family, which will be useful for evaluating the Kemeny constant.

\subsubsection{Eigenvalue spectrum}

Let $A_n$ and $D_n$ denote, respectively, the adjacency matrix and diagonal degree matrix of $H(q,n)$. The entry $A_n(i,j)$ of $A_n$ is defined as: $A_n(i,j)=1$ if nodes $i$ and $j$ are directly connected in $H(q,n)$, $A_n(i,j)=0$ otherwise. Then, the transition matrix of $H(q,n)$, denoted by $T_n$, is defined as $T_n=(D_n)^{-1}A_n$. Thus, the entry of $T_n$ is $T_n (i,j)= A_n(i,j)/d_i(n)$.

We now begin to compute the eigenvalues of $T_n$. Since $T_n$ is asymmetric, we introduce a real symmetric matrix $P_n$ similar to $T_n$ by
\begin{equation}\label{Mat03}
P_n=(D_n)^{-\frac{1}{2}} A_n (D_n)^{-\frac{1}{2}}=(D_n)^{\frac{1}{2}}T_n (D_n)^{-\frac{1}{2}}.
\end{equation}
Thus, we reduce the problem of determining eigenvalues of $T_n$ to calculating those of $P_n$, with entries $P_n(i,j)=\frac{A_n(i,j)}{\sqrt{d_i(n)}\sqrt{d_j(n)}}$.

Next, we apply the decimation method~\cite{DoAlBeKa83,CoKa92,ZhHuShCh12,WuZh12} to compute the eigenvalues and their multiplicities of $P_n$. Let us consider the eigenvalues of matrix $P_{n+1}$. For this purpose, we use $\alpha$ to denote the set of nodes belonging to $H(q,n)$, and $\beta$ the set of nodes created at iteration $n+1$. By definition, $P_{n+1}$ has the following block form:
\begin{equation}\label{T2}
P_{n+1}=\left[\begin{array}{cccc}
P_{\alpha,\alpha} & P_{\alpha, \beta} \\
P_{\beta,\alpha} & P_{\beta, \beta}
\end{array}
\right]
=\left[\begin{array}{cccc}
0 & P_{\alpha, \beta} \\
P_{\beta,\alpha} & 0
\end{array}
\right],
\end{equation}
where we have used the fact that $P_{\alpha,\alpha}$ and $P_{\beta, \beta}$ are two zero matrices with orders being $N_n\times N_n$ and $(N_{n+1}-N_n)\times (N_{n+1}-N_n)$, respectively. The reason for $P_{\alpha,\alpha}$ and $P_{\beta, \beta}$ being zero matrices lies in the iterations and insertion constraint in the construction process of networks, as shown in Fig.~\ref{cons}.

Suppose $\lambda_{i}(n+1)$ is an eigenvalue of $P_{n+1}$, and $u=(u_{\alpha},u_{\beta})^\top$ is its associated
eigenvector, where $u_{\alpha}$ and $u_{\beta}$ correspond to nodes in $\alpha$ and $\beta$, respectively. Then, the
eigenvalue equation for matrix $P_{n+1}$ has the following block form:
\begin{equation}\label{T1}
\left[\begin{array}{cccc}
0 & P_{\alpha, \beta} \\
P_{\beta,\alpha} & 0
\end{array}
\right]
\left[\begin{array}{cccc}
 u_{\alpha} \\
 u_{\beta}
\end{array}
\right]={\lambda}_{i}(n+1) \left[\begin{array}{cccc}
 u_{\alpha} \\
 u_{\beta}
\end{array}
\right]\,,
\end{equation}
which can be rewritten as
\begin{equation}\label{T3}
P_{\alpha, \beta}u_{\beta}=\lambda_{i}(n+1)u_{\alpha},
\end{equation}
\begin{equation}\label{T4}
P_{\beta,\alpha}u_{\alpha}=\lambda_{i}(n+1)u_{\beta}.
\end{equation}
In the case of $\lambda_{i}(n+1)\neq 0$, Eq.~(\ref{T4}) gives rise to
\begin{equation}\label{T5}
u_{\beta}=\frac{1}{\lambda_{i}(n+1)}P_{\beta,\alpha}u_{\alpha}\,.
\end{equation}
Inserting Eq.~(\ref{T5})
into Eq.~(\ref{T3}) yields
\begin{equation}\label{T6}
\frac{1}{\lambda_{i}(n+1)}P_{\alpha,\beta}P_{\beta,\alpha}u_{\alpha}=\lambda_{i}(n+1)u_{\alpha}.
\end{equation}
Thus, the problem of determining the eigenvalues
for matrix $P_{n+1}$ of order $N_{n+1} \times N_{n+1}$ is reduced
to calculating the eigenvalues of matrix $P_{\alpha,\beta}P_{\beta,\alpha}$ with a smaller
order of $N_{n} \times N_{n}$.

In Appendix~\ref{AppA}, we prove that
\begin{equation}\label{T7}
P_{\alpha,\beta}P_{\beta,\alpha}=\frac{1}{2}I_{n}+\frac{1}{2}P_{n},
\end{equation}
where $I_{n}$ is the identity matrix of order $N_{n} \times N_{n}$, same as that of $P_{n}$. Thus, we have related $P_{\alpha,\beta}P_{\beta,\alpha}$ to $P_{n}$, which allows to express the eigenvalues of matrix $P_{n+1}$ in terms of those of $P_{n}$.

Plugging Eq.~(\ref{T7}) into Eq.~(\ref{T6}) gives
\begin{equation}\label{T10}
\frac{1}{\lambda_{i}(n+1)}\left(\frac{1}{2}I_{n}+\frac{1}{2}P_{n}\right)u_{\alpha}
=\lambda_{i}(n+1)u_{\alpha},
\end{equation}
that is,
\begin{equation}\label{T12}
P_n u_{\alpha}=\left\{2[\lambda_{i}(n+1)]^2-1\right\}u_{\alpha}.
\end{equation}
Hence, if $\lambda_{i}(n)$ is an eigenvalue of $P_{n}$ corresponding to the eigenvector $u_{\alpha}$, then Eq.~(\ref{T12}) implies
\begin{equation}\label{T13}
\lambda_{i}(n)=2[\lambda_{i}(n+1)]^2-1.
\end{equation}
Solving the quadratic equation~(\ref{T13}) in  variable $\lambda_{i}(n+1)$ gives two roots:
\begin{equation}\label{T16}
\lambda_{i,1}(n+1)= \sqrt{\frac{\lambda_{i}(n)+1}{2}}, \quad \lambda_{i,2}(n+1)=-\sqrt{\frac{\lambda_{i}(n)+1}{2}}\,,
\end{equation}
which relate $\lambda_{i}(n+1)$ to $\lambda_{i}(n)$, with each $\lambda_{i}(n)$ producing two eigenvalues of
$P_{n+1}$.

\subsubsection{Multiplicities of eigenvalues}

To determine the multiplicities of eigenvalues, we first calculate numerically the eigenvalues for those networks of small sizes. For $n =0$, the eigenvalues of $P_{0}$ are $1$ and $-1$; while for $n =1$, the eigenvalues of matrix $P_{1}$ corresponding to $H_{q,1}$ are $1$, $-1$, and $0$ with multiplicity  $q$. For $n\geq 2$, we find that the eigenvalue spectrum displays the following properties: (i) Eigenvalues $1$ and $-1$ are present at all generations, both having a single degeneracy. (ii) All eigenvalues of generation $n_{i}$ will always exist at its subsequent generation $n_{i}+1$, and all new eigenvalues at generation $n_{i}+1$ are just those generated via Eq.~(\ref{T16}) by substituting $\lambda_{i}(n)$ with $\lambda_{i}(n_i)$ that are newly added to generation $n_{i}$; moreover, each new eigenvalue keeps the degeneracy of its father.  (iii) Except for $1$ and $-1$, all other eigenvalues are derived from generated $0$. Thus, all that is left to determine are the degeneracy of $0$ and the multiplicities of its offsprings, based on property (ii).

Let $D^{\rm mul}_n(\lambda)$ denote the multiplicity of eigenvalue $\lambda$ of matrix $P_n$. We now find the number
of eigenvalue 0 of $P_n$. Let $r(M)$ denote the rank of matrix $M$. Then, the degeneracy of eigenvalue 0 of
$P_{n+1}$ is
\begin{equation}\label{N0}
D^{\rm mul}_{n+1}(\lambda=0)= N_{n+1}-r(P_{n+1})\,.
\end{equation}
In Appendix~\ref{AppB}, we derive the exact expression for the degeneracy of eigenvalue 0 of $P_n$ as
\begin{equation}\label{Mult_0}
D^{\rm mul}_n(\lambda=0)=\begin{cases}
0, &n=0, \\
\frac{(q-1)(2q)^{n}+q}{2q-1}, &n \geq 1.
\end{cases}
\end{equation}
Since every eigenvalue in $P_n$ keeps the degeneracy of its father, the multiplicity of each first-generation
descendants of eigenvalue 0 is $\frac{(q-1)(2q)^{n-1}+q}{2q-1}$, the multiplicity of each second-generation
descendants of eigenvalue 0 is $\frac{(q-1)(2q)^{n-2}+q}{2q-1}$, and so on. Thus, for $P_n$ ($n\geq 1$), the total
number of eigenvalue 0 and all of its descendants is
\begin{eqnarray}\label{N6}
N_n^{\rm seed}(0)&=&\sum_{i=1}^{n}
\left[\frac{(q-1)(2q)^{i}+q}{2q-1}\times 2^{n-i}\right] \nonumber \\
&=&\frac{q(2q)^{n}-q}{2q-1}.
\end{eqnarray}
Summing up the multiplicities of eigenvalues obtained above, we have
\begin{eqnarray}\label{N7}
N_n^{\rm seed}(0)+D^{\rm mul}_n(-1)+D^{\rm mul}_n(1)=\frac{q(2q)^{n}+3q-2}{2q-1}=N_n.\nonumber \\
\end{eqnarray}
Thus, we have found all the eigenvalues of $P_n$.

Thus far, we have given a recursive expression for eigenvalues of the transition matrix for   $H(q,n)$, as well as the multiplicity of each eigenvalue. In fact, for the special network family $H(q,n)$ considered here, their eigenvalues can be expressed in an explicit way, as shown in Appendix~\ref{AppC}. These exact expressions contain more concrete information about eigenvalues than that of the recursive relation provided by Eq.~(\ref{T16}).

\subsection{The Kemeny constant} \label{Kemeny}

We proceed to use the obtained eigenvalues and their multiplicities to determine the Kemeny constant for the fractal scale-free networks $H(q,n)$.

Let $F_{ij}(n)$ denote the MFPT from node $i$ to node $j$ in $H(q,n)$. Let $\pi=(\pi_1, \pi_2,\ldots, \pi_{N_n})^\top$ represent the stationary distribution~\cite{Lo96,AlFi99} for random walks on $H(q,n)$. It is easy to verify that $\pi_i=d_i(n)/(2E_n)$. Then, the Kemeny constant for $H(q,n)$, denoted by $F_n$, is given by
\begin{equation}\label{eig01}
F_n=\sum_{j=1}^{N_{n}}\pi_j\,F_{ij}(n)=\sum_{i=2}^{N_{n}}\frac{1}{1-\lambda_{i}(n)},
\end{equation}
where we have assumed that $\lambda_{1}(n)=1$.

Let $L_n$ be the normalized Laplacian matrix~\cite{Ch97,ChLuVu03} of $H(q,n)$, defined as
$L_n=I_n-P_n=I_n-(D_n)^{\frac{1}{2}} T_n (D_n)^{-\frac{1}{2}}$. Let $\sigma_1$, $\sigma_2$, $\sigma_3$, $\cdots$,
$\sigma_{N_n}$ be the $N_n$ real non-negative eigenvalues of matrix $L_n$, rearranged as
$0=\sigma_1 <\sigma_2 \leq \sigma_3 \leq \ldots \leq \sigma_{N_n} = 2$.
Note that for any $i$, $\sigma_{i}(n)=1-\lambda_{i}(n)$.
Then,
\begin{equation}\label{eig02}
F_n=\sum_{i=2}^{N_{n}}\frac{1}{\sigma_{i}(n)}.
\end{equation}
Next, we explicitly evaluate this sum.

Let $\Omega_n$ be the set of all the $N_n-1$ nonzero eigenvalues of matrix $L_n$, i.e., $\Omega_n=\{ \sigma_2(n), \sigma_3(n),\ldots,\sigma_{N_n}(n)\}$, in which the distinction of the elements has been ignored. It is clear that $\Omega_n$ ($n\geq 1$) includes $1$, $2$, and other eigenvalues generated by $1$. Then, $\Omega_n$ can be classified into three subsets represented by $\Omega_n^{(1)}$, $\Omega_n^{(2)}$ and $\Omega_n^{(3)}$, respectively. That is, $\Omega_n=\Omega_n^{(1)} \cup \Omega_n^{(2)} \cup \Omega_n^{(3)}$, where $\Omega_n^{(1)}$ consists of eigenvalue 1 with multiplicity $\frac{(q-1)(2q)^{n}+q}{2q-1}$, $\Omega_n^{(2)}$ contains only eigenvalue 2 with single degeneracy, and $\Omega_n^{(3)}$ includes those eigenvalues generated by 1. Obviously, $\sum_{\sigma_{i}(n) \in \Omega_n^{(1)}}\frac{1}{\sigma_{i}(n)}=\frac{(q-1)(2q)^{n}+q}{2q-1}$ and $\sum_{\sigma_{i}(n) \in \Omega_n^{(2)}}\frac{1}{\sigma_{i}(n)}=\frac{1}{2}$.

For each eigenvalue $\sigma_i(n-1)$ in $\Omega_{n-1}^{(3)}$, it generates two eigenvalues,  $\sigma_{i,1}(n)$ and $\sigma_{i,2}(n)$, belonging to $\Omega_{n}^{(3)}$, via the following equation:
\begin{equation}\label{eig03}
2[\sigma_i(n)]^2 - 4\sigma_i(n) + \sigma_i(n-1) = 0\,,
\end{equation}
which can be easily obtained from Eq.~(\ref{T13}). According to Vieta's formulas, we have
\begin{equation}\label{eig03x}
\sigma_{i,1}(n)+ \sigma_{i,2}(n)= 2
\end{equation}
and
\begin{equation}\label{eig03y}
\sigma_{i,1}(n)\times \sigma_{i,2}(n) = \frac{\sigma_i(n-1)}{2}\,.
\end{equation}
Then
\begin{equation}\label{eig04}
\frac{1}{\sigma_{i,1}(n)} + \frac{1}{\sigma_{i,1}(n)} = \frac{4}{\sigma_i(n-1)},
\end{equation}
which indicates that
\begin{eqnarray}\label{eig05}
\sum_{\sigma_{i}(n) \in \Omega_n^{(3)}}\frac{1}{\sigma_{i}(n)}&= & 4\,\sum_{\substack{\sigma_{i}(n-1) \in
\Omega_{n-1} \\ \sigma_{i}(n-1) \neq 2 }} \frac{1}{\sigma_{i}(n-1)}\nonumber \\
&= &4\left(F_{n-1}-\frac{1}{2}\right).
\end{eqnarray}
Thus, we have
\begin{equation}\label{eig06}
F_n = 4\,F_{n-1} +\frac{(q-1)(2q)^{n}+q}{2q-1}-\frac{3}{2}.
\end{equation}
Considering the initial condition $F_0=\frac{1}{2}$, Eq.~(\ref{eig06}) can be solved by distinguishing two cases:
$q=2$ and $q \geq 3$. For $q=2$,
\begin{equation}\label{eig07}
F_{n}=\frac{1}{18} \left(6n \times 4^{n}+4^{n+1}+ 5\right)\,;
\end{equation}
while for $q \geq 3$,
\begin{equation}\label{eig07x}
F_{n}=\frac{3(q-1)(2q)^{n+1}-2q(2q-1)4^{n}+4q^2-11q+6}{6(q-2)(2q-1)} \,.
\end{equation}

We now express $F_{n}$ as a function of the network size $N_{n}$. For the case of
$q=2$, we have $4^{n}=\frac{3}{2}N_n-2$ and $n=\log_4\big(\frac{3}{2}N_n-2\big)$, which can be easily obtained from
Eq.~(\ref{Nn}). Thus, Eq.~(\ref{eig07}) can be rewritten in terms of $N_{n}$ as
\begin{equation}\label{eig08}
F_n=  \left(\frac{1}{2}N_n-\frac{2}{3}\right)\log_4\left(\frac{3}{2}N_n-2\right)+6N_n-3\,.
\end{equation}
Similarly to the case of $q=2$, for $q \geq 3$ we can derive that $(2q)^{n}=\frac{(2q-1)N_n-3q+2}{q}$ and $4^{n}=\left[\frac{(2q-1)N_n-3q+2}{q}\right]^{\log_{2q}4}$. Inserting these relations into Eq.~(\ref{eig07x}) leads to
\begin{eqnarray}\label{eig08x}
F_n&=&\frac{q-1}{q-2}N_n+ \frac{22q^2-41q+18}{6(q-2)(2q-1)}\nonumber \\
&\quad& -\frac{q}{3(q-2)}\left[\frac{(2q-1)N_n-3q+2}{q}\right]^{\log_{2q}4} \,.
\end{eqnarray}

Equations~(\ref{eig08}) and~(\ref{eig08x}) show that, for large networks, i.e., $N_g\rightarrow \infty$, the leading terms in the Kemeny constant for $H(q,n)$ follow distinct scalings for the two cases of $q=2$ and $q \geq 3$. For the former case,
\begin{equation}\label{eig09}
F_{n} \sim N_{n}\ln N_{n}\,;
\end{equation}
while for the latter case,
\begin{equation}\label{eig09x}
F_{n} \sim N_{n}\,.
\end{equation}
It should be stressed that, for $q \geq 3$, the linear growth of the Kemeny constant for sparse $H(q,n)$ is similar to that for dense complete graphs. In this sense, we say that the scale-free fractal networks $H(q,n)$ (for $q \geq 3$) are suboptimal networks.

Finally, we note that the validity of the above-obtained results for the Kemeny constant depends on that of the eigenvalues. In Appendix~\ref{AppD}, to verify that our computation on the eigenvalues of the transition matrix is correct, we compute the number of spanning trees in $H(q,n)$ using the connection between spanning trees and the eigenvalues of the normalized Laplacian matrix, which reproduces previously obtained results for the number of spanning trees in $H(q,n)$~\cite{ZhLiWuZo11}. This indeed validates our computations.

\section{Conclusions}

To conclude, by applying a different method from before, we have shown that among all networks, complete graphs are optimal in the sense that they have the minimum Kemeny constant, which scales linearly with the network size. However, due to the limitations of cost and other factors, complete graphs are not typical in the real world, where most systems are described by sparse networks displaying simultaneously scale-free and fractal properties. To explore networks having such properties of real systems but with a small Kemeny constant, we have also studied random walks on a scale-free fractal network family, characterized by a positive integer parameter. We have derived explicit formulas for all eigenvalues and their multiplicities of the transition matrix for such networks. Using these results, we have further determined the eigentime identity, which also displays a linear growth with the network size as complete graphs. At last, to test our results on the spectra of the transition matrix for the scale-free fractal  networks, we have computed the number of their spanning trees, which is consistent with those previously obtained results by means of a different method. Our work may have important implications on the design of complex networks with efficient navigation.

\begin{acknowledgments}
This work was supported by the National Natural Science Foundation
of China under Grant Nos. 61074119 and 11275049, and the Hong Kong Research Grants
Council under the GRF Grant CityU 1114/11E.
\end{acknowledgments}

\appendix

\section{Proof of equation~(\ref{T7})}\label{AppA}

In order to prove $P_{\alpha,\beta}P_{\beta,\alpha}=\frac{1}{2}I_{n}+\frac{1}{2}P_{n}$, it  suffices to show that their corresponding entries are equal to each other. For simplicity, let  $Q_n=\frac{1}{2}I_{n}+\frac{1}{2}P_{n}$, the entries of which are: $Q_n(i,i)=\frac{1}{2}$ for $i=j$ and $Q_n(i,j)=\frac{1}{2}P_n(i,j)$ otherwise. Let $R_n=P_{\alpha,\beta}P_{\beta,\alpha}$, whose entries $R_n(i,j)$ can be determined as follows. Note that $P_{\alpha,\beta}$ can be written as
\begin{equation} \label{App1}
P_{\alpha,\beta} =
\left(
\begin{array}{c}
M_1^\top\\
M_2^\top\\
\vdots\\
M^\top_{N_{n}}
\end{array}
\right),
\end{equation}
where each $M_i$ is a column vector of order $N_{n+1}-N_{n}$. Since $P_{\alpha,\beta}=P_{\beta,\alpha}^\top$, we have $P_{\beta,\alpha}=\left( M_1~M_2~ \ldots M_{N_{n}}\right)$. Then, the entry $R_n(i,j)=M_i^\top M_j$ of $R_n$ can be determined by distinguishes two cases, as follows.

If $i=j$, then the diagonal element of $R_n$ is
\begin{eqnarray}\label{App2}
R_n(i,i)&=& M_i^\top M_i \nonumber \\
&=& \displaystyle \sum_{k \in \beta} P_{n+1}(i,k)P_{n+1}(k,i)\nonumber \\
&=&\displaystyle \sum_{k \in \beta} \frac{A_{n+1}(i,k)}{d_i(n+1)d_k(n+1)} \nonumber \\
&=& \frac{1}{d_i(n+1)}\sum_{k \in \beta, \, i\thicksim k  } \frac{1}{2} \nonumber \\
&=&\frac{1}{2}\,,
\end{eqnarray}
where $i\thicksim k$
stands for that nodes $i$ and $k$ are adjacent in $H_{q,n+1}$ and the fact $d_k(n+1)=2$ was used.

If $i \neq j$, the non-diagonal element of $R_n$ is
\begin{eqnarray}\label{App3}
&\quad&R_n(i,j)= M_i^\top M_j \nonumber \\
&=&  \sum_{k \in \beta} P_{n+1}(i,k)P_{n+1}(k,j) \nonumber \\
&=& \sum_{\substack{A_{n+1}(i,k) = 1 \\ A_{n+1}(k,j) = 1 }} \frac{A_{n}(i,j)}{d_k(n+1)\sqrt{d_i(n+1)d_j(n+1)}}
\nonumber \\
&=& \frac{A_{n}(i,j)}{2\sqrt{d_i(n)d_j(n)}}=\frac{1}{2}P_n(i,j)=Q_n(i,j),
\end{eqnarray}
where the relations $d_i(n+1)=q\,d_i(n)$ and $d_j(n+1)=q\,d_j(n)$ have been used. Thus, Eq.~(\ref{T7}) is proved.

\section{Derivation of equation~(\ref{Mult_0})}\label{AppB}

Equation~(\ref{N0}) shows that in order to determine $D^{\rm mul}_{n+1}(\lambda=0)$, we can alternatively compute $r(P_{n+1})$. Obviously,
$r(P_{n+1})=r(P_{\alpha,\beta})+r(P_{\beta,\alpha})=2r(P_{\beta,\alpha})$, where
$r(P_{\alpha,\beta})=r(P_{\beta,\alpha})$ is used.

We next determine $r(P_{\beta,\alpha})$. As defined in Appendix~\ref{AppA}, $P_{\beta,\alpha}=\left( M_1~M_2~ \cdots M_{N_{n}}\right)$, where $M_{i}=(M_{1,i},M_{2,i},\cdots,M_{N_{n+1}-N_{n},i})^\top$. We first show that $M_1$, $M_2$, $\ldots$, and $M_{N_{n}}$ are linearly dependent. To this end, we label one node in $H(q,0)$ by $1$. Then, all nodes in $H(q,n+1)$ can be categorized into $2^{n+1}+1$ classes $C_i$ ($i=0, 1,\ldots,2^{n+1}$), according to the shortest  distance between any of them and node $1$. Let $d(x,y)$ denote the shortest distance from node $x$ to node $y$ in $H(q,n+1)$, and $C_i$ be the  set of nodes  defined by $C_i =\left\{u|d(1,u) = i \right\}$. Then, the set $\alpha$ of old nodes belonging to $H(q,n+1)$ is $\alpha = \displaystyle \bigcup_{i = 0}^{2^{n}}C_{2i}$, and set $\beta$ of new nodes born at generation $n+1$ is $\beta=\displaystyle \bigcup_{i = 1}^{2^{n} - 1}C_{2i-1}$. It is not difficult to verify that
\begin{equation}
\sum_{\substack{u = 1 \\ u \in \alpha }}^{N_{n}} \left[(-1)^{d(1,u)/2} M_{u}\right]=0.
\end{equation}
Therefore, column vectors $M_1$, $M_2$, $\ldots$, and $M_{N_{n}}$ are linearly dependent.

We next demonstrate that the $N_{n}-1$ vectors $M_2$, $M_3$, $\ldots$, and $M_{N_{n}}$ are linearly independent. Let
\begin{equation}
v=(v_2,v_{3},\ldots,v_{N_{n+1}-N_{n}})^\top=\sum_{\substack{i = 2 \\ i \in \alpha }}^{N_{n}} k_i M_{i}.
\end{equation}
Suppose that $v=0$, and we will prove that for an arbitrary $k_i$, $k_i=0$ always holds. Note that for a node $w \in \beta$, it has two neighbors, denoted by $w_{\rm pre}$ and $w_{\rm suc}$: $w_{\rm pre}$ is the neighbor near by node $1$, while $w_{\rm suc}$ is the other neighbor farther from node $1$. If $w \in C_1$, then $v_w=k_{w_{\rm pre}}+k_{w_{\rm suc}}$. Since, in this case, $k_{w_{\rm pre}}$ is just node $1$,  $v_w=0$ leads to $k_{w_{\rm suc}}=0$. Let's consider a node $j \in C_2$. By construction, there exists a node $w \in C_1$ such as $w_{\rm suc}=j$. Thus, if $v_j=0$, $k_{j}=0$ holds for $j \in C_2$. In a similar way, we can prove that for any node $i \in \alpha$ ($i \neq 1$), if $v=(0,0,\ldots,0)$, then $k_{i}=0$. In this way, we have proved that $M_1$, $M_2$, $\ldots$, and $M_{N_{n}}$ are linearly independent. Therefore, $r(P_{\beta,\alpha})=N_n-1=\frac{q(2q)^{n}+q-1}{2q-1}$. Inserting this formula into Eq.~(\ref{N0}), we can obtain the degeneracy of eigenvalue 0 of $P_n$, given by Eq.~(\ref{Mult_0}).

\section{Explicit expression of eigenvalues of $P_n$}\label{AppC}

Notice that all the eigenvalues of $P_n$ lie between $-1$ and $1$.
Let $\lambda_n$ be an eigenvalue of $P_n$ ($n \geq 0$). Then, one can assume that $\lambda_{n}=\cos(\theta_n)$ with
$0 \leq \theta_{n} \leq \pi$. Thus, from Eq.~(\ref{T13}), we have
\begin{equation}\label{EigExp01}
\cos^2(\theta_n)=\cos^2(\theta_{n-1}/2)\,,
\end{equation}
which implies that
\begin{equation}\label{EigExp02}
\cos(\theta_n) = \pm \cos(\theta_{n-1}/2).
\end{equation}
Under the constraint of $0 \leq \theta_{n} \leq \pi$, Eq.~(\ref{EigExp02}) means
\begin{equation}\label{EigExp03}
\theta_n = \theta_{n-1}/2 ~~~\mbox{or}~~~
\theta_n = \pi - \theta_{n-1}/2\,,
\end{equation}
which enable us to express the eigenvalues for $P_n$ ($n \geq 0$) in an explicit way as follows.

We use an ordered pair $(\lambda_n, D^{\rm mul}_{\lambda_n})$ to represent an eigenvalue $\lambda_n=\cos \theta_{n}$ of matrix $P_n$ and its corresponding multiplicity $D^{\rm mul}_{\lambda_n}$. Let $\Phi_n$ denote the set of eigenvalues of matrix $P_n$. Since $\Phi_n$ includes eigenvalues $-1$, $1$, $0$, and those derived from $0$,  $\Phi_n$ can be classified into $n+1$ non-crossing subsets, denoted by $\Phi_{n,0}$, $\Phi_{n,1}$, $\Phi_{n,2}$, $\dots$, $\Phi_{n,n-1}$, and $\Phi_{n,n}$, respectively. $\Phi_{n,0}$ contains eigenvalues $-1$ and $1$ with single degeneracy; $\Phi_{n,1}$ contains all eigenvalues $0$; $\Phi_{n,2}$ contains all child eigenvalues directly generated from $\Phi_{n-1,1}$; $\Phi_{n,3}$ contains all child eigenvalues directly generated from $\Phi_{n-1,2}$; and so on. Let $\chi_i=\frac{(q-1)(2q)^{n+1-i}+q}{2q-1}$.  Then, $\Phi_n$ can be explicitly expressed as follows:
\begin{equation}
\Phi_n = \bigcup_{i = 0}^{n} \Phi_{n,i}\,,
\end{equation}
where
\begin{equation}
\Phi_{n,0} = \left\{\left(\cos 0,1 \right), \left(\cos \pi,1 \right)\right\}
\end{equation}
and
\begin{equation}
\Phi_{n,i} = \bigcup_{k = 1}^{2^{i-1}} \left\{\left(\cos \frac{2k - 1}{2^i}\pi,\chi_i \right)\right\}\,,
\end{equation}
for $1 \leq i \leq n$.

For example, for $n=3$, we have
\begin{equation}
\Phi_3 = \Phi_{3,0} \bigcup \Phi_{3,1} \bigcup \Phi_{3,2} \bigcup \Phi_{3,3}\,, \nonumber
\end{equation}
where
\begin{equation}
\Phi_{3,0}=\left\{\left(\cos 0,1 \right), \left(\cos \pi,1 \right)\right\}\,, \nonumber
\end{equation}
\begin{equation}
\Phi_{3,1}=\left\{\left(\cos \frac{\pi}{2}, \frac{(q-1)(2q)^{3}+q}{2q-1}  \right) \right \}\,, \nonumber
\end{equation}
\begin{small}
\begin{equation}
\Phi_{3,2}=\left\{\left(\cos \frac{\pi}{4}, \frac{(q-1)(2q)^{2}+q}{2q-1}  \right),\left(\cos \frac{3\pi}{4},
\frac{(q-1)(2q)^{2}+q}{2q-1}  \right)  \right \}\,, \nonumber
\end{equation}
\end{small}
and
\begin{small}
\begin{equation}
\Phi_{3,3}=\left\{\left(\cos \frac{\pi}{8}, q  \right),\left(\cos \frac{3\pi}{8}, q  \right),\left(\cos
\frac{5\pi}{8}, q  \right),\left(\cos \frac{7\pi}{8}, q  \right) \right \}. \nonumber
\end{equation}
\end{small}

\section{Number of spanning trees in $H(q,n)$}\label{AppD}

Let $N_{\rm st}(q,n)$ denote the number of spanning trees in $H(q,n)$. According to the well-known results~\cite{Ch97,ChZh07}, the number of spanning trees, $N_{\rm st}(q,n)$, can be determined by $N_n-1$ nonzero eigenvalues of the normalized Laplacian matrix for $H(q,n)$, as
\begin{equation}\label{ST01}
N_{\rm st}(q,n)=\frac{\displaystyle{\prod_{i=1}^{N_{n}}
d_i(n)}\prod_{i=2}^{N_{n}}\sigma_i(n)}{\displaystyle {\sum_{i=1}^{N_{n}}d_i(n)}}\,.
\end{equation}

Let $\Theta_n$, $\Delta_n$ and $\Lambda_n$ denote $\sum_{i=1}^{N_{n}}d_i(n)$, $\prod_{i=1}^{N_{n}} d_i(n)$ and
$\prod_{i=2}^{N_{n}}\sigma_i(n)$, respectively. Then, we have
\begin{equation}\label{ST01x}
N_{\rm st}(q,n)=\frac{\Delta_n \times \Lambda_n}{\Theta_n}\,
\end{equation}
and
\begin{equation}\label{ST01y}
N_{\rm st}(q,n-1)=\frac{\Delta_{n-1} \times \Lambda_{n-1}}{\Theta_{n-1}}\,.
\end{equation}

We next determine the recursive relation between $N_{\rm st}(q,n)$ and $N_{\rm st}(q,n-1)$. We first determine the recursive relations for $\Theta_n$, $\Delta_n$ and $\Lambda_n$. It is evident that
\begin{equation}\label{ST02}
\Theta_n=\sum_{i=1}^{N_{n}}d_i(n)=2E_n=2(2q)^{n}=2q \Theta_{n-1}.
\end{equation}
Moreover, according to the obtained results in the main text, it is easy to derive the following recursive
relations:
\begin{equation}\label{ST03}
\Delta_n =q^{N_{n-1}} \Delta_{n-1}\times 2^{qE_{n-1}}=q^{N_{n-1}}\times 2^{q(2q)^{n-1}}\Delta_{n-1}
\end{equation}
and
\begin{equation}\label{ST04}
\Lambda_n =\left(\frac{1}{2}\right)^{N_{n-1}-2}\Lambda_{n-1}.
\end{equation}

Combining the above-obtained relations, we  obtain the recursive relation for $N_{\rm st}(q,n)$ as
\begin{equation}\label{ST01z}
N_{\rm st}(q,n)=q^{N_{n-1}-1}2^{q(2q)^{n-1}-N_{n-1}+1} N_{\rm st}(q,n-1)\,.
\end{equation}
Using $N_{\rm st}(q,0)=1$, we can now solve Eq.~(\ref{ST01z}) to obtain the exact expression for the number of spanning trees in $H(q,n)$ as
\begin{widetext}
\begin{equation}\label{ST05}
N_{\rm
st}(q,n)=2^{\frac{2q-nq-2q^2+2nq^2-2^{n+1}q^{n+1}+2^{n+1}q^{n+2}}{(2q-1)^2}-n}q^{\frac{-q+nq-2nq^2+2^{n}q^{n+1}}{(2q-1)^2}+n}\,,
\end{equation}
\end{widetext}
which is in complete agreement with that obtained in~\cite{ZhLiWuZo11} yet from a different approach, implying that the technique and process of our computation on the eigenvalues and their degeneracies for $H(q,n)$ are correct.

\bibliography{aipsamp}

\end{document}